%% file: main.tex
\documentclass{article}
\usepackage{multirow}
\usepackage{styles,times}
\usepackage{lipsum}
\usepackage{multirow}
\usepackage{booktabs}
\usepackage{multirow}
\usepackage{multicol}
\usepackage{adjustbox}
\usepackage{tcolorbox}
\usepackage{tabularx}
\usepackage{tikz}
\usepackage{graphicx}

\usepackage{subfig}
\usepackage{caption}

\input{math_commands.tex}

\usepackage{hyperref}
\usepackage{url}
\usepackage{lineno}
\usepackage{array, makecell}
\usepackage{algorithm}
\usepackage{algpseudocode}
\usepackage[numbers]{natbib}
\usepackage{geometry}
\usepackage{amssymb}
\usepackage{chngcntr}
\usepackage{subcaption}
\usepackage{enumitem}
\usepackage{overpic}

\usepackage{float} 
\title{\textsc{Improving Reliability of Machine Learned Interatomic Potentials With Physics-Informed Pretraining}}

\author{Qianyu Zheng$^{\dag}$, Victor Fung$^{\dag\text{\textsection}}$ \\
{\normalsize $^\dag$College of Computing, Georgia Institute of Technology, Atlanta, GA, USA} \\
{\normalsize $^\ddag$School of Computational Science and Engineering, Georgia Institute of Technology, GA, USA}\\
{\normalsize $^{\text{\textsection}}$Corresponding authors: 
\texttt{victorfung@gatech.edu}}
}
\date{}

\newcolumntype{Y}{>{\centering\arraybackslash}X}

\iclrfinalcopy 
\begin{document}

\maketitle

\begin{abstract}

Machine learned interatomic potentials (MLIPs) have emerged as powerful tools for molecular dynamics (MD) simulations with their competitive accuracy and computational efficiency. However, MLIPs are often observed to exhibit un-physical behavior when encountering configurations which deviate significantly from their training data distribution, leading to simulation instabilities and unreliable dynamics, thus limiting the reliability of MLIPs for materials simulations. We present a physics-informed pretraining strategy that leverages simple physical potentials which can improve the robustness and stability of graph-based MLIPs for MD simulations. We demonstrate this approach by deploying a pretraining-finetuning pipeline where MLIPs are initially pretrained on data labelled with embedded atom model potentials and subsequently finetuned on the quantum mechanical ground truth data. 
By evaluating across three diverse material systems (phosphorus, silica, and a subset of Materials Project) and three representative MLIP architectures (CGCNN, M3GNet, and TorchMD-NET), we find that this physics-informed pretraining consistently improves both prediction accuracy as well as stability in MD compared to the baselines. 


\end{abstract}

\section{Introduction}\label{sec:intro}


Molecular Dynamics (MD) simulations are a powerful method used in computational physical science research which captures the evolution of atomic configurations over time and enables the investigations of a broad range of molecular, biological, and material behaviors and properties.\cite{MD-elastic, Fracture-Toughness-MD, Plastic-MD, Solid-Solid-MD, Melting-MD, Reaction-Pathway-MD, Surface-Adsorption-MD, Active-Site-MD1, Active-Site-MD2}. However, a central challenge lies in extending the spatial and temporal limits of these simulations while maintaining appreciable accuracy, which requires models that can evaluate potential energy surfaces with both high efficiency and fidelity. \textit{Ab initio} methods, such as density functional theory (DFT) \cite{DFT, Electron-Density-DFT}, are capable of highly accurate evaluations of the potential energy surface, providing high-fidelity results with no parameterization needed. While they remain the gold standard for these simulations, the demanding computational scaling of these methods with respect to system size severely limits their practical applicability in large system size or long-timescale molecular dynamics. This motivates the search for alternative strategies that retain near-\textit{ab initio} accuracy while significantly reducing computational cost. Machine Learning Interatomic Potentials (MLIPs) have now attracted significant attention as a promising solution to this problem. By learning latent representations of atoms in various chemical environments without requiring specific knowledge about chemical bonds or interactions, MLIPs provide competitive accuracy and computational efficiency superior to \textit{ab initio} methods in atomistic modeling \cite{MLFF-overview}, which makes them particularly valuable for large-scale MD simulations.

With recent advances in deep learning, end-to-end neural networks have become the dominant architecture found in MLIPs. Early neural network potentials like the Behler-Parrinello Neural Network (BP-NN) \cite{BP} and ANAKIN-ME (ANI) \cite{ANI} demonstrated the capacity for learning potential energy surfaces directly from large datasets, however they still relied on careful construction of structure representations to serve as features for the models. These were followed by the development of graph neural network (GNN) architectures that represent atomic systems as graphs, with notable early examples including SchNet \cite{SchNet}, DimeNet \cite{DimeNet}, M3GNet \cite{M3GNet}, CHGNet \cite{CHGNet}, and ALIGNN-FF \cite{ALIGNN-FF}. To better capture physical laws and improve data efficiency, subsequent approaches integrated E(3)-equivariance into their architectures. Equivariant models like NequIP \cite{NequIP}, Allegro \cite{Allegro}, SevenNet \cite{SevenNet}, and EquiformerV2 \cite{Equiformer} use tensor-based representations and equivariant convolutions to preserve rotational and translational symmetries. More recent developments have demonstrated the potential of universal interatomic potentials trained on diverse, large-scale datasets: examples include MACE-MP-0 \cite{MACE-MP-0}, Mattersim \cite{MatterSim}, Orb \cite{Orb} for inorganic crystals, MACE-OFF23 \cite{MACE-OFF23} for organic molecules, AIMNet2 \cite{AIMNet2} for charged organic systems.

However, despite MLIPs' impressive performance in predicting the potential energy surfaces of complex atomistic systems as seen in recent benchmarks \cite{Matbench-Discovery}, MLIPs exhibit concerning unphysical behaviors when encountering structural configurations that deviate significantly from their training data distribution \cite{MD-OOD, Transferability-Study}. These out-of-distribution scenarios are particularly problematic during MD, as these simulations naturally explore configurational space through thermal fluctuations, structural deformations, and chemical reactions, inevitably encountering regions far from the equilibrium and which are not adequately represented in the training datasets \cite{DIRECT-Sampling, Active-Learning-Config, Systematic-Softening}. As noted in recent studies, aligning the training data distribution to any configuration space explored during test-time MD simulations may be an impossible task \cite{Active-Learning-Config}, and we can not expect MLIPs to demonstrate high physical reliability in out-of-distribution structural configurations that drift significantly from the structures in the training set \cite{Systematic-Softening}. This reliability crisis represents a manifestation of the broader Out-of-Distribution (OOD) problem in machine learning \cite{General-OOD}, where models trained on finite datasets struggle to generalize beyond their training domain, but with particularly severe consequences in MD simulations where unphysical predictions can propagate and compound over time \cite{MD-Error-Propagation}, ultimately leading to unrecoverable errors in trajectories and unscientific MD trajectories \cite{Transferability-Study}.

Several strategies have emerged to address MLIP robustness: active learning, physical constraints, and pretraining on large datasets. Active learning uses intelligent data selection to continuously expand the training dataset to mitigate the OOD problem. Uncertainty-driven approaches by Kulichenko et al. \cite{Uncertainty-AL} and Zhang et al. \cite{Active-lr-2} guide data acquisition, while Van der Oord et al. \cite{HAL-AL} use Hyperactive Learning to automate continuous integration of new structures into the training datasets. Matsumura et al. \cite{Active-lr-small-dist} target unstable structures in expanding the dataset. To overcome the significant computational cost induced by iterative retraining and expensive, Sanjeev et al. \cite{StaBlE} reduce costs by sampling failing trajectories to retrain on system observables such as RDF instead of labels obtained by first-order methods. Integrating physical constraints into neural networks provide an alternative pathway to promote physical output from MLIPs. Pun et al. \cite{PINN} directly combines traditional physics-based potentials as extra terms into neural networks, Ibayashi et al. \cite{Allegro-Legato} use Sharpness-Aware Minimization for robustness, and Fu et al. \cite{SN-Reg} apply Spectral Norm Regularization while incorporating Morse potential constraints. Lastly, pretraining on large datasets increases the structural diversity exposed to the MLIP during training and thus enhances its OOD performance. Maheshwari et al. \cite{simple-pretraining} show OC20 pretraining enables extended stable simulations, Qi et al. \cite{DIRECT-AL} developed DIRECT sampling for the Materials Project for creating pretraining datasets with data efficiency, and Takamoto et al. \cite{Taylor-Expansion-Label} use Taylor expansion in data augmentation to create a large weakly labeled dataset to train the model. Other approaches include classic force field pretraining \cite{TIP3P-Pretrain}, Gardner et al.'s \cite{Distillation-With-Low-Fidelity} data augmentation with rattle-relax-repeat and subsequent knowledge distillation, and Cui et al.'s \cite{IPIP-Pretrain} iterative framework combining pseudo-labeling with feedback from MD simulation.

Despite the aforementioned progress, several limitations exist which could hinder their widespread adoption for building robust MLIPs for MD simulations. Active learning methods, while effective at improving accuracy in targeted regions, suffer from prohibitive computational costs due to iterative retraining cycles and the need for additional expensive \textit{ab initio} calculations during simulation. Regularization and physics-informed methods, although computationally more tractable, face challenges in striking an optimal balance between incorporating sufficient physical constraints and maintaining the flexibility necessary for capturing the potential energy surface. 
These limitations motivate the need for alternative approaches that can improve MLIP reliability without excessive computational overhead while providing robust mechanisms for preventing unphysical behaviors in MD simulations.



In this paper, we present our work covering three key aspects. First, we introduce how traditional empirical potentials can benefit machine learning interatomic potentials (MLIPs) by injecting physical knowledge through a pretraining-finetuning workflow. Second, we design and implement two novel benchmarking metrics that evaluate MLIPs by detecting unphysical behaviors in the MLIP-generated molecular dynamics trajectories. Third, through experiments with three different MLIPs and benchmark datasets using our proposed benchmark suite, we demonstrate the competitiveness of our method in improving the reliability of MLIPs for empowering molecular dynamics simulations and perform further analysis of the physical knowledge transfer process to further show the theoretical feasibility of our framework.

\section{Methodology}\label{sec:method}

\subsection{Machine Learned Interatomic Potentials}\label{sec:MLIPs}

We evaluate our physics-informed pretraining framework and trajectory physicality benchmarking suite using three representative MLIP architectures that embody different design philosophies in the field: Crystal Graph Convolutional Networks (CGCNN) \cite{CGCNN}, M3GNet \cite{M3GNet} and TorchMD-NET \cite{TorchMD}. CGCNN is a classic graph neural network model for materials, where atomic structures are encoded as crystal graphs with atoms as nodes and bonds as edges. The model employs graph convolutional operations between atoms to iteratively update atomic representations by aggregating information from neighboring atoms. This model is invariant to translations, rotations, and permutations of the atoms in the system. \cite{CGCNN} M3GNet represents an improved example of an invariant machine learning interatomic potential that incorporates three-body interactions to achieve high accuracy across diverse materials systems. The model encodes atomic structures in a way that incorporates both distance and angular information, which enables it to better learn complex many-body interactions. M3GNet demonstrates exceptional transferability across different chemical compositions and structural motifs and is thus effective for materials discovery and property prediction tasks. \cite{M3GNet} TorchMD-NET is an equivariant transformer model which represents a further jump in complexity over the previous two examples. This model is equivariant to translations and rotations of the atoms and incorporates an attention based mechanism to further improve model expressivity. \cite{TorchMD} The selection of these three architectures allows us to assess the transferability of our proposed methods across different MLIP design paradigms and evaluate how our physics-informed pretraining and benchmarking approaches perform across distinct representation learning frameworks.

\subsubsection{Datasets} \label{sec:datasets}

We evaluate our methodology across three datasets of different chemical systems: Phosphorous, Silica and MPtrj subset. The Phosphorous dataset comprises 4,798 structural configurations of phosphorus across multiple phases and morphologies. It includes GAP-RSS configurations from iterative random sampling, liquid structures (both molecular $P_4$ and network phases), two-dimensional materials like phosphorene with defects, bulk crystals at ambient and high pressures up to 100 GPa, and isolated molecular fragments. The structural diversity and inclusion of both stable and metastable configurations across the wide chemical space provides a stringent benchmark for evaluating MLIP transferability and accuracy. All structures are labeled with total energy and forces per atom by DFT computations \cite{P-dataset}. The Silica dataset consists of 2864 amorphous and crystalline silica structures with liquid, glass, and crystalline phases, including distorted crystalline configurations, isolated dimers, and melt-quench trajectories. The dataset is constructed by iteratively expanding a reference database generated through multiple cycles of GAP fitting and molecular dynamics simulations and is subsequently labeled with single-point DFT computation of total energy, forces per atom and cell stress \cite{Silica-dataset}. The MPtrj subset dataset is a curated dataset sampled from the Materials Project Trajectories (MPtrj) dataset \cite{MPtrj}, which contains relaxation trajectories for tens of thousands of crystal structures computed with DFT. Our subset consists of 4965 structures extracted from the MPtrj dataset that contain exclusively lithium (Li), manganese (Mn), oxygen (O), and phosphorus (P) as constituent elements. The MPtrj subset provides a more complicated case for MLIP performance on materials systems, as it contains structures with more types of atoms than the silica and phosphorous dataset, varying degrees of structural disorder and different oxidation states of metalic elements. The presence of multiple ionic species with distinct bonding characteristics, from the highly ionic interactions to non-bonding iteractions, creates a challenging environment for testing the transferability and stability of machine learning force fields.

\subsection{Physics-Informed Pretraining Framework}

\subsubsection{Leveraging Physics-based Empirical Potentials}

\begin{figure}[h!]
\centering
\includegraphics[width=1.0\textwidth]{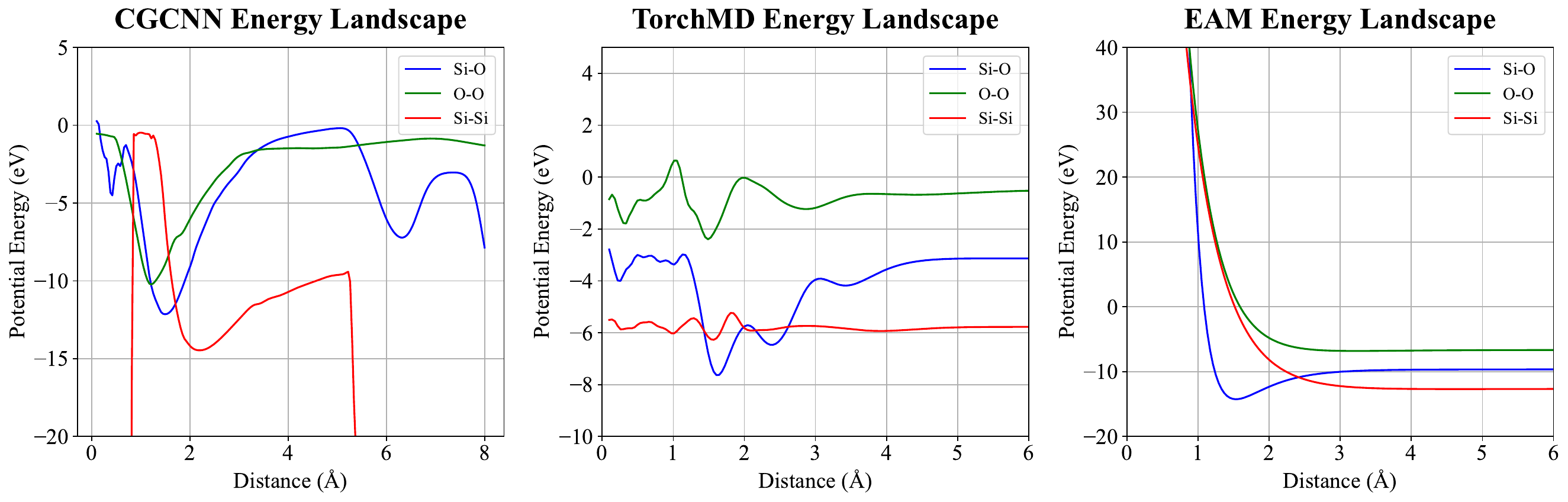}
\caption{Predicted energy landscapes from MLIP CGCNN and TorchMD, and the EAM potential for energy prediction of three types of interactions in the Silica dataset. Clear unphysical patterns are observed in both MLIPs below 1\r{A} and above 6\r{A} for CGCNN.}
\label{fig:physical_vs_unphysical}
\end{figure}

Traditional empirical interatomic potentials, despite their inferior accuracy compared to quantum mechanical methods because of their fixed functional forms, can encode fundamental physical principles. These potentials inherently capture physical behaviors that are often challenging for purely data-driven MLIPs to learn consistently from finite training datasets. Most critically, with their functional forms, traditional interatomic potentials generally satisfy constraints at limits of coordinates in that they predict energy $E(r = 0) \rightarrow \infty$ and $E(r \rightarrow \infty) = 0$ \cite{Traditional-Pot-Characteristics}, though different potentials achieve the constraints with different forms. Therefore, empirical potentials naturally incorporate short-range repulsive interactions that prevent unphysical atomic overlap, a behavior governed by the Pauli exclusion principle \cite{Pauli-Repulsion} that MLIPs often violate (as shown in Figure \ref{fig:physical_vs_unphysical}). This primarily stems from the characteristic of most existing material datasets in which structures close to their equilibrium states dominate, and lack structures containing pairs of atoms close to each other.

Additionally, traditional potentials generally produce well-shaped, smooth energy landscapes with physically meaningful curvature, which ensures stable force calculations and prevents the erratic energy fluctuations that can lead to simulation instabilities. By transferring this embedded physical knowledge into MLIPs, we hypothesize that the resulting models will exhibit enhanced robustness and improved adherence to fundamental physical constraints, particularly in unexplored regions of configurational space where pure machine learning approaches typically fail.

\subsubsection{Implementation of the Empirical Potential}

The Embedded Atom Method (EAM), as introduced in $\cite{EAM}$, is an empirical method for computing potential energy surfaces commonly for metallic systems. While modern MLIPs significantly outperform EAM in predicting forces and energies, the latter maintains several crucial advantages that make it particularly valuable for improving stability of MLIP-empowered MD simulation. EAM potentials typically contain only a few dozen parameters that need to be fitted, in contrast to MLIPs which typically require millions of parameters. In addition, its functional form ensures that they inherently respect physical principles. The robust and physically-grounded nature of EAM is a critical characteristic we hope to leverage in this project. With EAM, the total energy of a system is expressed as the sum of two distinct contributions: an embedding energy term that represents the energy required to place an atom into the electron density created by all surrounding atoms and a pairwise interaction term between each pair of atoms in the system. We implement this formulation \eqref{eqn:EAM} with an additional atomic energy term to the standard EAM formulation, which is explained later in this subsection. In the equation, $F_\alpha$ represents the energy embedding function \cite{EAM}, $\rho_\beta$ denotes the electron density function, and $V(r)$ represents the pairwise interaction function. In the following sections, we elaborate on the implementation details of the three components of the equation \eqref{eqn:EAM}.
\begin{equation}
    \label{eqn:EAM}
    E_{tot} = F_\alpha\left(\sum_{i\neq j}\rho_\beta(r_{ij})\right) + \frac{1}{2}\sum_{i\neq j} V(r_{ij}) + E_{atom}
\end{equation}

The determination of accurate energy embedding and electron density functions in EAM presents a fundamental challenge due to the complex quantum mechanical nature of electron density distributions in metals. While first-principles methods such as density functional theory (DFT) \cite{DFT, Electron-Density-DFT} could theoretically provide precise electron densities, their computational expense renders them impractical for large-scale applications. Following the approach suggested in \cite{EAM-exp-decay}, we employ multiple exponential decay functions to approximate the electron density term and adopt a simple quadratic form for the energy embedding function. Our electron density construction is formulated in equation \eqref{eqn:EAM-approx} with learnable parameters $A, \beta \in \mathbb{R}^{n_a \times n_b}$ and $ B, \rho_0 \in \mathbb{R}^{n_a}$, where $n_a$ represents the number of diatomic interaction types in the dataset and $n_b$ denotes the number of exponential basis functions employed.
\begin{equation}
    \label{eqn:EAM-approx}
    \begin{split}
        \rho_\beta(r_{ij}) &= \sum_{k=1}^{n_b} A_k \cdot \exp(-\beta_k \cdot r_{ij}) \\
        F_\alpha(\rho) &= B \cdot (\rho - \rho_0)^2 \text{, with } \rho = \sum_{j\neq i}\rho_\beta(r_{ij}) \\
    \end{split}
\end{equation}
For the second term representing the pairwise interaction function in equation \ref{eqn:EAM}, we employed the Morse potential. As originally introduced by \cite{Morse-Pot}, the Morse potential is an empirical interatomic potential function that describes the energy of a diatomic molecular bond as a function of interatomic distance. This potential offers significant advantages over simpler harmonic models by capturing both the anharmonicity of real chemical bonds and realistic dissociation behavior.
\begin{equation}
    \label{eqn:morse_e}
    V(r) = D_e\left(1 - e^{-a(r-r_e)}\right)^2
\end{equation}
For two atoms separated by distance $r$, the Morse potential energy $V(r)$ is given by equation \eqref{eqn:morse_e}. The learnable parameters are $D_e, a, r_e \in \mathbb{R}^{n_a}$, where $D_e$ represents the well depth (dissociation energy), $r_e$ denotes the equilibrium bond distance, and $a$ controls the width of the potential well. This potential formulation successfully captures both the anharmonic nature of real bonds and the dissociation behavior of molecules using a compact set of physically meaningful parameters.

At last, we add an additional atomic energy term $E_{atom}$ to the EAM formulation, which serves as a learnable reference energy that accounts for the isolated atom's electronic structure and sets a physically meaningful zero-point. This can improve the neural network's ability to predict absolute energies and formation energies across different chemical environments. It is implemented as the summation of atomic energy of each atom in the structure, a parameter depending on the element type of the atom learned together with other EAM parameters.

\begin{figure}[h!]
\centering
\includegraphics[width=1.0\textwidth]{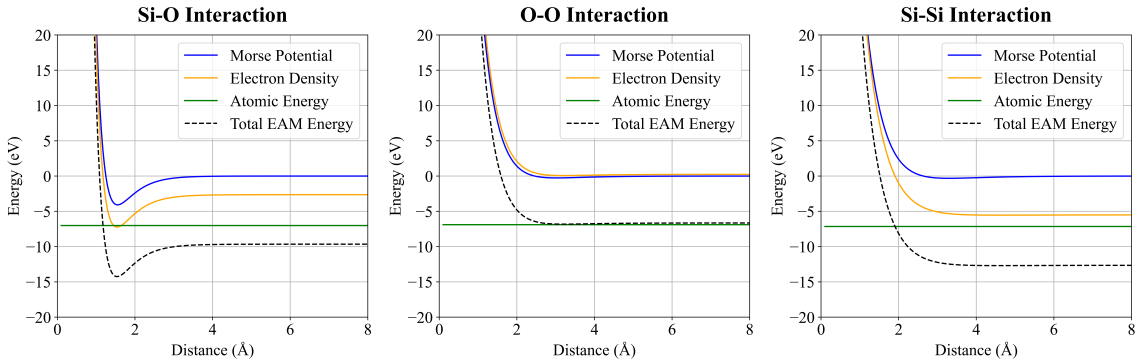}
\caption{Decomposition of the trained EAM potential for modeling three types of di-atomic interaction in the Silica dataset with individual energy components as a function of interatomic distance. The plots display the contribution of electron density (orange solid line), Morse pairwise interaction (blue solid line), atomic energy (green solid line) and total EAM energy (black dashed line) for the three primary atom pair types in silica: (left) Si-O interactions, (center) O-O interactions, and (right) Si-Si interactions. All energies are shown relative to the cutoff radius of 8.0 \r{A}.}
\label{Fig:silica-eam}
\end{figure}

We optimize all EAM potential parameters using backpropagation with the AdamW optimizer, minimizing the Mean Absolute Error (MAE) of energies, forces, and stress (EFS) by treating both electron density coefficients ($A$, $\beta$, $B$, $\rho_0$), Morse parameters ($D_e$, $a$, $r_e$) and per-element atomic energies as learnable variables. The equilibrium distance parameter $r_e$ is initialized to the sum of covalent radii for the corresponding atomic pair as a physically meaningful starting point, while all other parameters are initialized to unity. This initialization strategy ensures that the optimization begins from a chemically reasonable configuration while allowing the model to adaptively learn optimal parameter values through gradient-based training on the reference data.

Figure \ref{Fig:silica-eam} visualizes the decomposition of the trained EAM potential for the silica dataset across the three primary interaction types. The results demonstrate that while the Morse potential captures the fundamental pairwise bonding interactions, the electron density term provides corrections that enable the combined potential to achieve significantly lower EFS MAE compared to either component alone. The trained EAM potential successfully reproduces the expected chemical behavior of silica: strong attractive Si-O bonding with a potential minimum near 1.6 \r{A} (consistent with typical Si-O bond lengths in SiO$_2$ \cite{Si-O-bond}) and weaker O-O and Si-Si interactions. This decomposition illustrates how the EAM framework effectively provides a physically interpretable foundation for subsequent steps to improve the stability of the MD simulation enabled by MLIP.

\subsection{Effectiveness of the EAM Potential}

\begin{figure}[h!]
\centering
\includegraphics[width=1.0\textwidth]{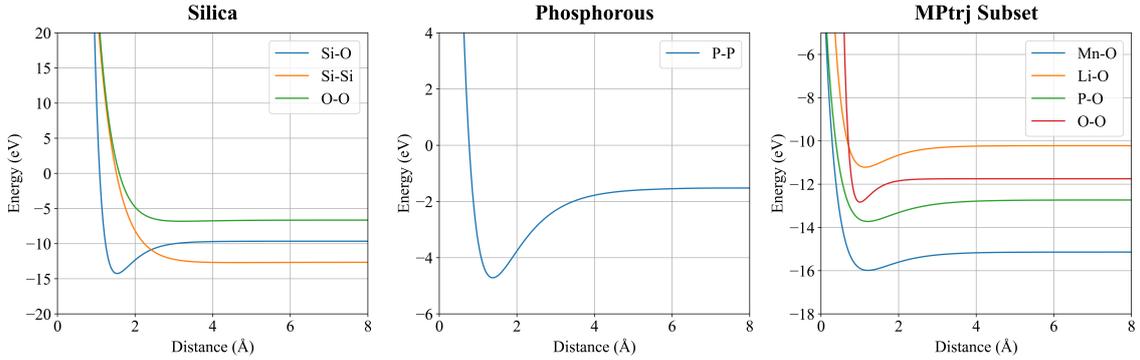}
\caption{EAM potential energy curves for key interatomic interactions across three benchmark datasets. The plots show the fitted EAM total energy as a function of interatomic distance for the most chemically significant atom pair interactions in each system: (left) Si-O, O-O and Si-Si interactions in the silica dataset, (center) P-P interactions in the phosphorus dataset, and (right) four representative interactions in the MPtrj subset.}
\label{Fig:EAM-fitting}
\end{figure}

The effectiveness of our fitted EAM potential can be demonstrated through its ability to capture interatomic interactions in material systems, as shown in Figure \ref{Fig:EAM-fitting}.
The fitted potential energy curves exhibit physically reasonable behavior for critical interactions with its appropriate short-range repulsive walls, well-defined energy minima and smooth asymptotic decay at large interatomic distances. Across the three benchmark datasets, silica, phosphorus, and the MPtrj subset, the EAM potential successfully reproduces the characteristic bonding patterns and energy scales of each chemical environment, with distinct curves representing different coordination states and local atomic environments. The energy profiles demonstrate that the EAM formulation provides sufficient flexibility to model diverse bonding types ranging from covalent Si-O interactions in silica to metallic bonding in transition metal systems. While the EAM potential may not achieve the quantitative accuracy of DFT calculations, as demonstrated by the imperfect well location for modeling interactions in the MPtrj Subset, it captures the fundamental physics necessary to generate physically meaningful pretraining data that guide the MLIP toward chemically sensible solutions and empirically promotes the physicality of the MD trajectories generated by the MLIPs based on the outcomes of our experiments.

\subsubsection{Physics-Informed Pretraining Workflow}

\begin{figure}[h!]
\centering
\includegraphics[width=0.8\textwidth]{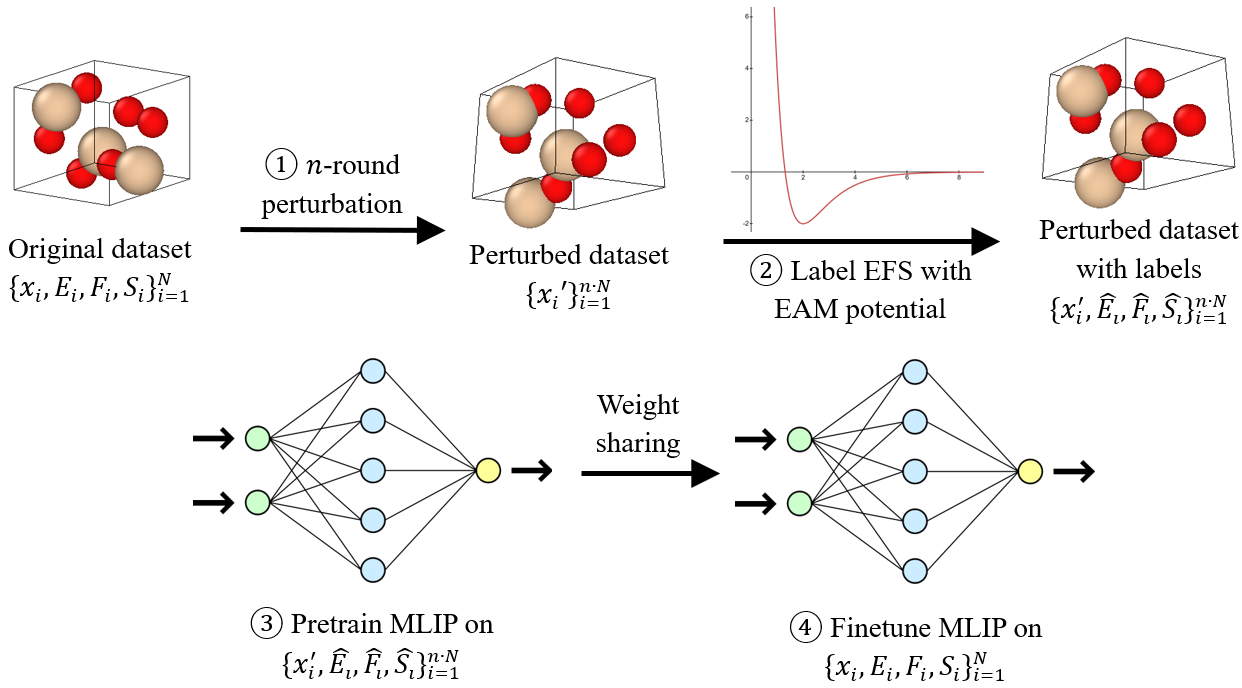}
\caption{Physics-informed pretraining workflow. The workflow begins with the construction of the pretraining dataset with the original quantum mechanical dataset. Through n-round perturbation following the two-step-perturbation algorithm, atomic positions are systematically modified to generate diverse structural configurations. The perturbed structures are then labeled using the trained EAM potential. The machine learning interatomic potential (MLIP) is first pretrained on this EAM-labeled dataset, then finetuned on the original quantum mechanical data through weight sharing.}
\label{Fig:workflow}
\end{figure}

We implement a systematic pretraining-finetuning pipeline that leverages the EAM potential to enhance MLIP robustness through structured exposure to physically meaningful perturbations. The complete methodology is illustrated in Figure \ref{Fig:workflow} and consists of two primary phases: pretraining dataset generation and subsequent model finetuning.

\textbf{Pretraining dataset construction}: we generate the pretraining dataset by systematically perturbing atomic positions in the original quantum mechanical reference structures. This augmentation strategy exposes the model to a broader range of configurational space while maintaining physical realism through EAM-based labeling.

For each reference structure with atomic positions $\mathbf{P} \in \mathbb{R}^{N_a \times 3}$, we generate perturbed configurations $\mathbf{P}' \in \mathbb{R}^{N_a \times 3}$ using a two-step perturbation algorithm that combines targeted atomic compression with random perturbation, as detailed in Algorithm \ref{alg:perturb}. This procedure first identifies close atomic pairs and systematically reduces their separation distances, followed by random perturbations of selected atoms, creating configurations that challenge the model with both coordination changes and general structural distortions.

Each perturbed structure is labeled using the trained EAM potential to compute energies, atomic forces, and stress tensors. This dual perturbation strategy generates configurations that systematically explore both coordination environment changes (through atomic compression) and general structural distortions (through random displacement), providing physically consistent reference data that enables the MLIP to learn robust interpolation behavior across diverse configurational landscapes.

\begin{algorithm}
\caption{Two-Step Atomic Position Perturbation}
\begin{algorithmic}[1]
\Require Atomic positions $\mathbf{P} \in \mathbb{R}^{N_a \times 3}$, maximum pull distance $d_{pull}$, maximum perturbation distance $d_{pert}$, number of close pairs $n_{close\_pairs}$, number of random sites $n_{sites}$
\State Compute pairwise distance matrix $\mathbf{D} \in \mathbb{R}^{N_a \times N_a}$
\State Identify $n_{close\_pairs}$ non-overlapping atom pairs with minimum interatomic distances
\For{each close pair $(i,j)$}
    \State $s \sim \mathcal{U}(0, d_{pull})$ \Comment{Sample compression strength}
    \State $\mathbf{d}_{ij} = \mathbf{P}_j - \mathbf{P}_i$
    \State $\mathbf{P}_i \gets \mathbf{P}_i + s \cdot \frac{\mathbf{d}_{ij}}{\|\mathbf{d}_{ij}\|_2}$ \Comment{Pull atoms closer}
    \State $\mathbf{P}_j \gets \mathbf{P}_j - s \cdot \frac{\mathbf{d}_{ij}}{\|\mathbf{d}_{ij}\|_2}$
\EndFor
\State Randomly select $n_{sites}$ atoms for random perturbation
\For{each selected atom $k$}
    \State $\mathbf{V}_k \sim \mathcal{N}(0,1)^{3}$, $\mathbf{V}_k \gets \mathbf{V}_k / \|\mathbf{V}_k\|_2$ \Comment{Random unit vector}
    \State $r \sim \mathcal{U}(0,d_{pert})$ \Comment{Sample perturbation magnitude}
    \State $\mathbf{P}_k \gets \mathbf{P}_k + r \cdot \mathbf{V}_k$ \Comment{Apply random displacement}
\EndFor
\State \Return $\mathbf{P}$ \Comment{Return perturbed positions}
\end{algorithmic}
\label{alg:perturb}
\end{algorithm}

\textbf{Pretraining and Finetuning}: the backbone MLIP is initially trained on the EAM-labeled augmented dataset. This pretraining phase establishes physically informed weight initialization and biases the model toward solutions that respect the underlying physics encoded in the EAM potential. Following pretraining, we employ transfer learning with weight sharing to finetune the model on the original quantum mechanical reference data. This finetuning process preserves the physical intuition learned during pretraining while achieving the accuracy of \textit{ab initio} methods. The resulting model serves as the final MLIP for subsequent MD simulation benchmarking and evaluation.

\subsection{Benchmarking Suite for Trajectory Physicality} \label{md-metrics}

\subsubsection{Molecular Dynamics Simulations} \label{sec:MD_intro}

Molecular dynamics simulations are used to evolve the state of an atomistic system with Newtonian equation of motion given the force acting on each atom at time $t$. With $T$ steps, the simulation produces a trajectory of positions $J = \{\mathbf{P}_t \in \mathbb{R}^{N\times 3}\}^T_{t=1}$, in which $\mathbf{P}_t$ represents the positions of atoms at timestep $t$. In this project, we primarily leverage NVT and NPT simulation with Langevin and Berendsen dynamics, respectively.

\subsubsection{Metrics for MD Trajectory Physicality} \label{sec:MD_metrics}

To quantitatively describe the quality of MLIP-generated MD trajectories, existing studies \cite{F-not-enough}\cite{StaBlE}\cite{Uncertainty-AL} typically rely on system observables such as radial distribution functions (RDF), diffusion coefficients, and vibrational density of states by comparing these properties between MLIP-generated trajectories and those produced by \textit{ab initio} methods. While these benchmarks successfully assess how well MLIPs can reproduce \textit{ab initio} reference data, they do not directly detect unphysical artifacts or violations of physical laws within the trajectories themselves. Therefore, in addition to these metrics, we introduced two structural metrics that detect structural abnormalities during the MD simulation exemplified by overlapping atoms and structural collapse.

Given a trajectory of $T$ timesteps $J = \{A_0, \dots, A_T\}$, where each $A_i$ represents the structural configuration at the $i^{th}$ timestep, our benchmarking suite consists of the following two physicality metrics:

\textbf{Overlapping Atoms}: This metric identifies catastrophic simulation failures where atoms approach unphysically close distances, violating fundamental principles of atomic structure. Such overlaps typically indicate inadequate representation of repulsive interactions in the MLIP.

A trajectory is classified as unphysical if, at any timestep, there exists a pair of atoms whose separation distance falls below a dynamically determined threshold:
\[\text{Trajectory $J$ is unphysical} \iff \exists k \in \{1, \dots, T\}, \min_{i \neq j \in \{0, \dots N\}}||(\mathbf{P}_{A_k})_i - (\mathbf{P}_{A_k})_j||_2 \leq \text{threshold} \]
To accommodate the compositional diversity across different material systems in our datasets, we define an adaptive threshold based on the initial equilibrium structure rather than employing a fixed universal cutoff:
\[\text{threshold} = \frac{1}{2}\min_{i \neq j \in \{0, \dots N\}}||(\mathbf{P}_{A_0})_i - (\mathbf{P}_{A_0})_j||_2\]
This adaptive approach ensures that the metric remains sensitive to structural violations while accounting for the inherent length scales of different chemical systems.

\textbf{Lindemann Index}: The Lindemann index quantifies thermal disorder and structural stability by measuring the magnitude of atomic positional fluctuations relative to equilibrium configurations \cite{L-Index-Def1}. Originally developed to characterize melting transitions, this metric serves as a sensitive indicator of excessive structural mobility that may signal simulation artifacts or unphysical dynamics.

For a structure containing $n$ atoms, the local Lindemann index $q_i$ for the $i^{th}$ atom quantifies the normalized root-mean-squared bond length fluctuation \cite{L-Index-Def2}, while the global index $q_{global}$ represents the system-wide average. The notation $\langle\cdot\rangle$ denotes temporal averaging over all timesteps:
\[
q_i = \frac{1}{N - 1}\sum_{j \neq i} \frac{\langle r^2_{ij}\rangle - \langle r_{ij}\rangle^2}{\langle r_{ij}\rangle},\quad q_{global} = \frac{1}{N}\sum_i q_i
\]
We employ the Lindemann index to detect excessive structural reorganization that exceeds physically reasonable thermal motion. A trajectory is deemed unphysical if any individual atom exhibits anomalously large fluctuations (indicating local instabilities) or if the entire system displays collective instability.

Given that experimental and theoretical studies establish melting behavior at Lindemann indices of 0.1-0.15 \cite{L-Index-threshold}, we conservatively set our thresholds at 0.25 for local indices and 0.2 for the global index to capture structural instabilities while avoiding false positives from legitimate thermal motion:
\[
\text{Trajectory $J$ is unphysical} \iff \exists i: q_i > 0.25 \text{ or } q_{global} > 0.2
\]

\section{Results}\label{sec:results}
In this section, we present experimental evidence to demonstrate the effectiveness of our approach for developing stable MLIPs for molecular dynamics simulations. We first evaluate the how pretraining on EAM-labeled data improves the performance of MLIPs in empowering physically reliable MD simulation with metrics proposed in section \ref{md-metrics} and how well the MLIP-empowered MD trajectories can resemble the DFT-empowered MD trajectories. We additionally evaluate the MLIPs' prediction accuracy for energy, forces, and stress (EFS) to demonstrate that our proposed method is minimally invasive and preserves the accuracy of structural property predictions.

\subsection{Experimental Setup}

To comprehensively evaluate the effectiveness of our physics-informed pretraining approach, we conduct experiments across three diverse datasets using three representative MLIP architectures: CGCNN, TorchMD-NET and M3GNet, as introduced in section \ref{sec:datasets} and \ref{sec:MLIPs}, respectively. Given that TorchMD-NET and M3GNet demonstrate superior baseline performance compared to CGCNN in terms of prediction accuracy, we implement a much stricter ``stress testing" protocol for TorchMD-NET and M3GNet by deliberately restricting the available training data, including the data used for EAM potential fitting. This constraint creates a more challenging scenario that better highlights the benefits of physics-informed pretraining under data-limited conditions.

\subsection{Physicality of MLIP-generated MD trajectories} \label{physicality_benchmark}

\begin{table}[htbp]
\centering
\renewcommand{\arraystretch}{1.3}
\begin{tabular}{|c|c|c|c|c|}
\hline
\textbf{Dataset} & \textbf{MD Setting} & \textbf{Setup} & \textbf{Atom Overlap} & \textbf{Lindemann Index} \\
\hline
\multirow{6}{*}{Silica (full)} & \multirow{3}{*}{NVT (2000K, 2fs)} & CGCNN & 5 & 2 \\
& & CGCNN + SAM & \textbf{4} & 4 \\
& & \textbf{CGCNN + Pretrain} & \textbf{4} & \textbf{0} \\
\cline{2-5}
& \multirow{3}{*}{NPT (2000K, 2fs)} & CGCNN & 9 & 7 \\
& & CGCNN + SAM & 8 & 11 \\
& & \textbf{CGCNN + Pretrain} & \textbf{5} & \textbf{2} \\
\hline
\multirow{10}{*}{Silica (10\%)} & \multirow{5}{*}{NVT (2000K, 2fs)} & TorchMD & 15 & 8 \\
& & TorchMD + SAM & 12 & 8 \\
& & \textbf{TorchMD + Pretrain} & \textbf{5} & \textbf{7} \\
\cline{3-5}
& & M3GNet & \textbf{2} & 3 \\
& & \textbf{M3GNet + Pretrain} & \textbf{2} & \textbf{2} \\
\cline{2-5}
& \multirow{5}{*}{NPT (2000K, 2fs)} & TorchMD & 14 & 7 \\
& & TorchMD + SAM & 8 & 8 \\
& & \textbf{TorchMD + Pretrain} & \textbf{5} & \textbf{6} \\
\cline{3-5}
& & \textbf{M3GNet} & \textbf{2} & \textbf{2} \\
& & \textbf{M3GNet + Pretrain} & \textbf{2} & \textbf{2} \\
\hline
\multirow{3}{*}{Phosphorous (full)} & \multirow{3}{*}{NVT (1250K, 1fs)} & CGCNN & 35 & 70 \\
& & CGCNN + SAM & 34 & \textbf{64} \\
& & CGCNN + Pretrain & \textbf{22} & 65 \\
\cline{2-5}
\hline
\multirow{5}{*}{Phosphorous (5\%)} & \multirow{5}{*}{NVT (1250K, 1fs)} & TorchMD  & 9 & 42 \\
& & TorchMD + SAM & 25 & 57 \\
& & \textbf{TorchMD + Pretrain} & \textbf{6} & \textbf{24} \\
\cline{3-5}
& & M3GNet & \textbf{0} & 43 \\
& & \textbf{M3GNet + Pretrain} & \textbf{0} & \textbf{4} \\
\hline
\multirow{6}{*}{MPtrj Subset (full)} & \multirow{3}{*}{NVT (500K, 1fs)} & CGCNN & 89 & 39 \\
& & CGCNN + SAM & 7 & 5 \\
& & \textbf{CGCNN + Pretrain} & \textbf{2} & \textbf{4} \\
\cline{2-5}
& \multirow{3}{*}{NPT (250K, 1fs)} & CGCNN$^{\dag}$ & * & * \\
& & CGCNN + SAM & 8 & 9 \\
& & \textbf{CGCNN + Pretrain} & \textbf{4} & \textbf{6} \\
\hline
\multirow{10}{*}{MPtrj Subset (10\%)} & \multirow{5}{*}{NVT (1000K, 1fs)} & TorchMD & 70 & 98 \\
& & TorchMD + SAM & 49 & 97 \\
& & \textbf{TorchMD + Pretrain} & \textbf{11} & \textbf{40} \\
\cline{3-5}
& & M3GNet & 100 & 100 \\
& & \textbf{M3GNet + Pretrain} & \textbf{0} & \textbf{19} \\
\cline{2-5}
& \multirow{5}{*}{NPT (750K, 1fs)} & TorchMD & 84 & 98 \\
& & TorchMD + SAM & 28 & 72 \\
& & \textbf{TorchMD + Pretrain} & \textbf{1} & \textbf{16} \\
\cline{3-5}
& & M3GNet & 100 & 100 \\
& & \textbf{M3GNet + Pretrain} & \textbf{0} & \textbf{82} \\
\hline
\end{tabular}
\caption{Trajectory stability assessment for hybrid MLIPs across diverse datasets and simulation conditions. All molecular dynamics simulations were conducted for 5000 timesteps on 100 randomly sampled structures from each dataset's test set. Values represent the number of trajectories (out of 100) that exhibit unphysical behavior as detected by atom overlap and Lindemann index criteria. NVT and NPT simulations are performed with Langevin and Berendsen dynamics, respectively, as introduced in section \ref{sec:MD_intro}, and implemented with the Atomic Simulation Environment (ASE) package. Due to the absense of stress labels in the Phosphorous dataset and the requirement of stress tensor of NPT simulations, NPT experiments were not performed on the Phosphorous dataset.}
\label{tab:md_results}
\end{table}

To evaluate our physics-informed pretraining approach under realistic simulation conditions, we conducted comprehensive molecular dynamics benchmarking across all three datasets. For each dataset, we randomly sampled 100 structures from the respective test sets and performed 5000-step MD simulations using different thermodynamic ensembles and temperature conditions. The MD simulations serve as a test of model robustness, as they require the MLIP to maintain physical consistency over extended trajectories while exploring configurational space beyond the training distribution. To show the effectiveness of our method, we benchmark the proposed method against the baseline MLIP training and the MLIP trained with SAM optimization \cite{Allegro-Legato}. We assessed simulation stability using the two metrics introduced in \ref{sec:MD_metrics}: overlapping atoms and the Lindemann index. All results are presented in table \ref{tab:md_results}.

Our physics-informed pretraining approach demonstrates substantial improvements in molecular dynamics simulation stability across datasets, MLIP architecture and thermodynamic ensembles, as quantified by reduced atom overlap incidents and Lindemann index violations during extended simulations. The ability to achieve zero atom overlaps in multiple test cases indicates that our approach successfully enforces short-term repulsion in MLIPs with the injected physical knowledge from the EAM potential, which improves MLIPs reliability for extended molecular dynamics applications.

\subsection{Capability of MLIPs to recover DFT MD trajectories}

\begin{figure*}
\centering
\includegraphics[width=\textwidth]{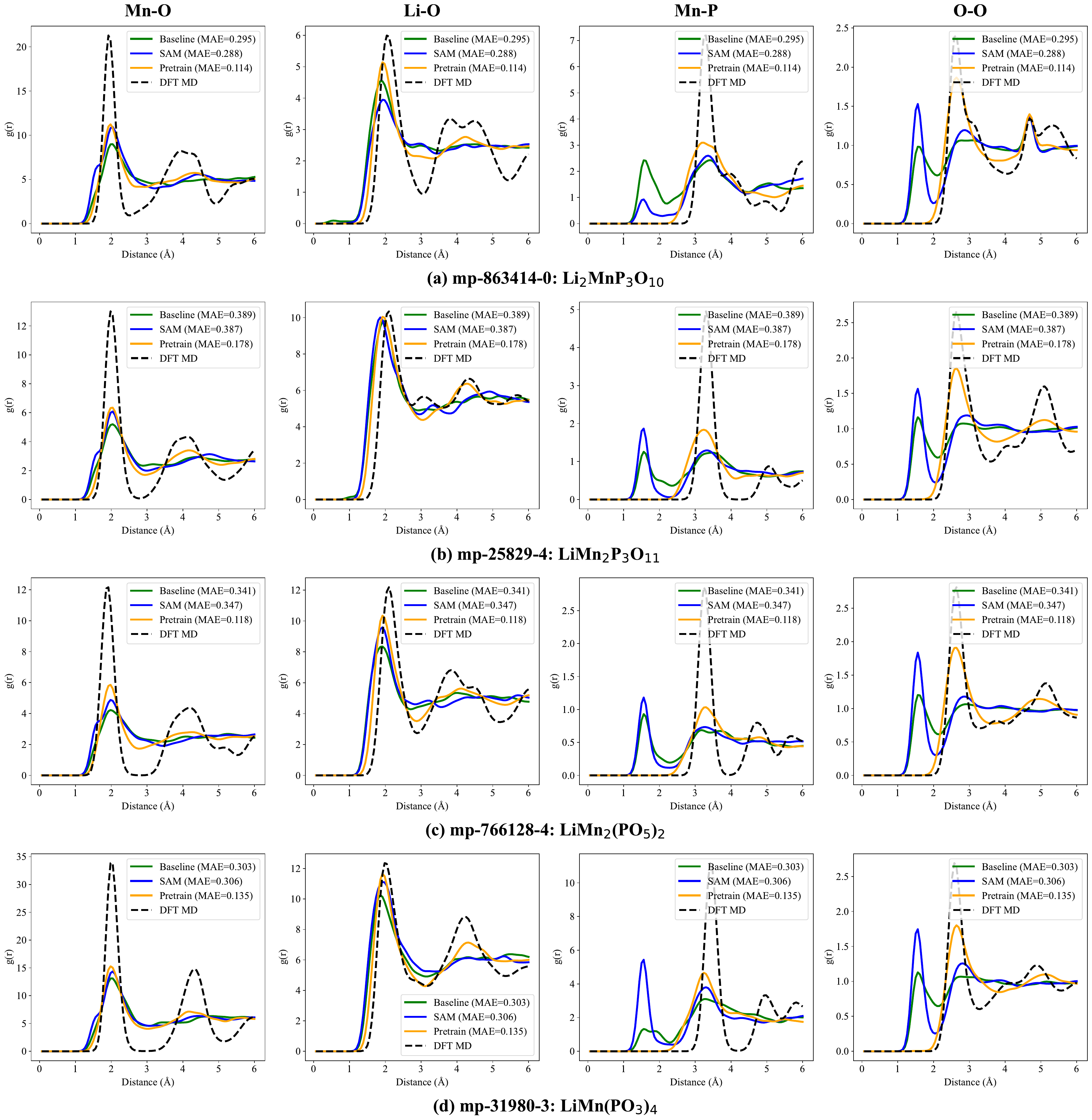}
\caption{Radial distribution function comparison between DFT molecular dynamics and MLIP predictions for four representative MPtrj Subset structures. The plots show RDFs for four atomic pair interactions (Mn-O, Li-O, Mn-P, O-O) across four selected structures: (a) mp-863414-0, (b) mp-25829-4, (c) mp-766128-4, and (d) mp-31980-3. Black dashed lines represent reference DFT MD trajectories, while colored lines show MLIP predictions using different training approaches: orange curves show our proposed pretraining method, blue curves show SAM optimization results, and green curves represent the baseline vanilla TorchMD model.}
\label{Fig:rdf}
\end{figure*}

Besides evaluating the physicality of MLIP-generated trajectories through the aforementioned two benchmarking metrics, we additionally assess MLIP performance by comparing structural properties with reference DFT molecular dynamics simulations. To establish this comparison, we conducted DFT Langevin NVT simulations on ten representative structures selected from the MPtrj Subset dataset, using identical initial configurations and simulation conditions as employed in the corresponding MLIP NVT simulations. We utilize Radial Distribution Functions (RDF) as a quantitative metric for structural comparison, following the approach established in \cite{F-not-enough}. RDFs serve as system observables that capture the statistical distribution of interatomic distances and provide insight into structural ordering. By comparing the RDFs derived from DFT MD trajectories with those from MLIP MD simulations, we evaluate the MLIPs' capability to accurately reproduce the structural dynamics and correlation patterns observed in the quantum mechanical reference simulations. Figure \ref{Fig:rdf} presents the RDF analysis for four representative atomic pair types: Li-O, Mn-O, O-O, and Mn-P. The first two pairs (Li-O and Mn-O) represent typical coordination environments with strong chemical bonding interactions commonly found in the MPtrj Subset structures, while the latter two pairs (O-O and Mn-P) correspond to non-bonded interactions that provide insight into medium-range structural correlations.

The RDF comparison results demonstrate that our physics-informed pretraining methodology provides substantial improvements in modeling diverse atomic interactions across the four representative MPtrj Subset structures with notable enhancements for both bonded and non-bonded pairs. The RDF of MD trajectory generated by the physically informed pretrained TorchMD model (orange lines) consistently achieves superior agreement with DFT reference trajectories (black dashed lines) compared to both the SAM optimization method (blue lines) and the baseline vanilla TorchMD model (green lines) across all interaction types. For the primary bonded interactions (Mn-O and Li-O), our method reproduces the characteristic first coordination shell peaks around 1.5-2.5 \r{A} and maintains proper peak intensities and shapes that closely match the DFT reference data.

Most significantly, for the challenging non-bonded interactions (Mn-P and O-O), the pretraining approach successfully eliminates the unphysical artifacts observed in both alternative methods, which exhibit dubious short-distance peaks in regions where strong repulsion should dominate and RDF values should approach zero. These improvements are physically meaningful because the presence of anomalous short-distance peaks indicates that MLIPs are predicting chemically implausible atomic configurations with unrealistic close contacts. The enhanced physicality of our pretraining-based MLIPs in capturing proper repulsive behavior at short distances is consistent with the reduced occurrences of atomic overlap abnormalities documented in our trajectory physicality benchmarks (Section \ref{physicality_benchmark}). The consistent performance across structures further validates the robustness and transferability of our physics-informed pretraining approach for complex oxide materials.

\subsection{Prediction Accuracy of MLIPs}

Furthermore, we demonstrate that the proposed method introduces minimal harm and often provides additional benefits to the accuracy of MLIPs. We employ energy, force, and stress (EFS) Mean Absolute Error (MAE) (energy and forces MAE for the Phosphorous dataset) on the test set of each dataset as benchmarking metrics, as they directly quantify MLIPs' capability to predict material properties during MD simulations. Table \ref{tab:model_performance} presents a comprehensive comparison of prediction errors with the baseline MLIP, MLIP trained with SAM optimization \cite{Allegro-Legato} and MLIP with the proposed EAM-assisted pretraining.

\begin{table}[htbp]
\centering
\renewcommand{\arraystretch}{1.3}
\begin{tabular}{|c|c|c|c|c|c|}
\hline
\textbf{Dataset} & \textbf{Setup} & \textbf{E Error} & \textbf{F Error} & \textbf{S Error} & \textbf{Total Error} \\
\hline
\multirow{3}{*}{Silica (full)} & CGCNN & 29.062 & 0.208 & 0.00290 & 10.796 \\
& CGCNN + SAM & 27.722 & 0.216 & 0.00306 & 11.071 \\
& \textbf{CGCNN + Pretrain} & 23.871 & 0.202 & 0.002209 & \textbf{10.450} \\
\hline
\multirow{5}{*}{Silica (10\%)} & TorchMD & 8.157 & 0.174 & 0.00117 & 6.010 \\
& \textbf{TorchMD + SAM} & 5.071 & 0.107 & 0.000876 & \textbf{5.460} \\
& TorchMD + Pretrain & 7.820 & 0.113 & 0.000374 & 5.797 \\
\cline{2-6}
& M3GNet & 3.252 & 0.114 & 0.00441 & 5.967 \\
& \textbf{M3GNet + Pretrain} & 4.069 & 0.112 & 0.00286 & \textbf{5.781} \\
\hline
\multirow{3}{*}{Phosphorus (full)} & CGCNN & 9.976 & 0.252 & / & 12.677 \\
& CGCNN + SAM & 10.778 & 0.265 & / & 13.338 \\
& \textbf{CGCNN + Pretrain} & 9.476 & 0.232 & / & \textbf{11.709} \\
\hline
\multirow{5}{*}{Phosphorus (5\%)} & \textbf{TorchMD} & 16.301 & 0.201 & / & \textbf{10.222} \\
& TorchMD + SAM & 7.439 & 0.233 & / & 11.712 \\
& TorchMD + Pretrain & 22.196 & 0.203 & / & 10.331 \\
\cline{2-6}
& M3GNet & 8.170 & 0.257 & / & 12.931 \\
& \textbf{M3GNet + Pretrain} & 23.033 & 0.248 & / & \textbf{12.629} \\
\hline
\multirow{3}{*}{MPtrj (full)} & CGCNN & 1.485 & 0.0914 & 0.00191 & 4.679 \\
& CGCNN + SAM & 3.518 & 0.0847 & 0.00180 & 4.359 \\
& \textbf{CGCNN + Pretrain} & 2.575 & 0.0782 & 0.00167 & \textbf{4.017} \\
\hline
\multirow{5}{*}{MPtrj (10\%)} & TorchMD & 6.139 & 0.0931 & 0.00184 & 4.806 \\
& \textbf{TorchMD + SAM} & 4.910 & 0.0736 & 0.00150 & \textbf{3.803} \\
& TorchMD + Pretrain & 4.945 & 0.0861 & 0.00179 & 4.443 \\
\cline{2-6}
& M3GNet & 3.065 & 0.0854 & 0.00231 & 4.416 \\
& \textbf{M3GNet + Pretrain} &  6.568 & 0.0652 & 0.00171 & \textbf{3.412} \\
\hline
\end{tabular}
\caption{Prediction accuracy comparison for hybrid MLIPs combining MLIPs with EAM potentials. Energy errors are reported as mean absolute error (MAE) in eV, force errors in eV/\r{A}, and stress errors in GPa. The total error represents a weighted combination of energy, force, and stress MAEs using weights 0.01:50:50, respectively. (The Phosphorous dataset has energy and force weights of 0.01:50 due to the absence of stress labels.) Dataset percentages indicate the fraction of available training data used for model development. The ``Setup" column includes three experimental setups: the baseline MLIP, the SAM-optimized MLIP and the finetuned MLIP with the proposed pretraining strategy. The experimental setups with the lowest total error are highlighted.}
\label{tab:model_performance}
\end{table}

The results demonstrate that our physics-informed pretraining approach is largely non-intrusive, in that the finetuned MLIP maintains competitive prediction accuracy with modest but consistent improvements over baseline models. Across all benchmark datasets, the pretraining methodology does not compromise the fundamental predictive capabilities of the underlying MLIP architectures, with total error metrics remaining within reasonable ranges compared to their respective baselines.

\subsection{Knowledge Transfer from EAM potential to MLIP}

\begin{figure}[h!]
\centering
\includegraphics[width=1.0\textwidth]{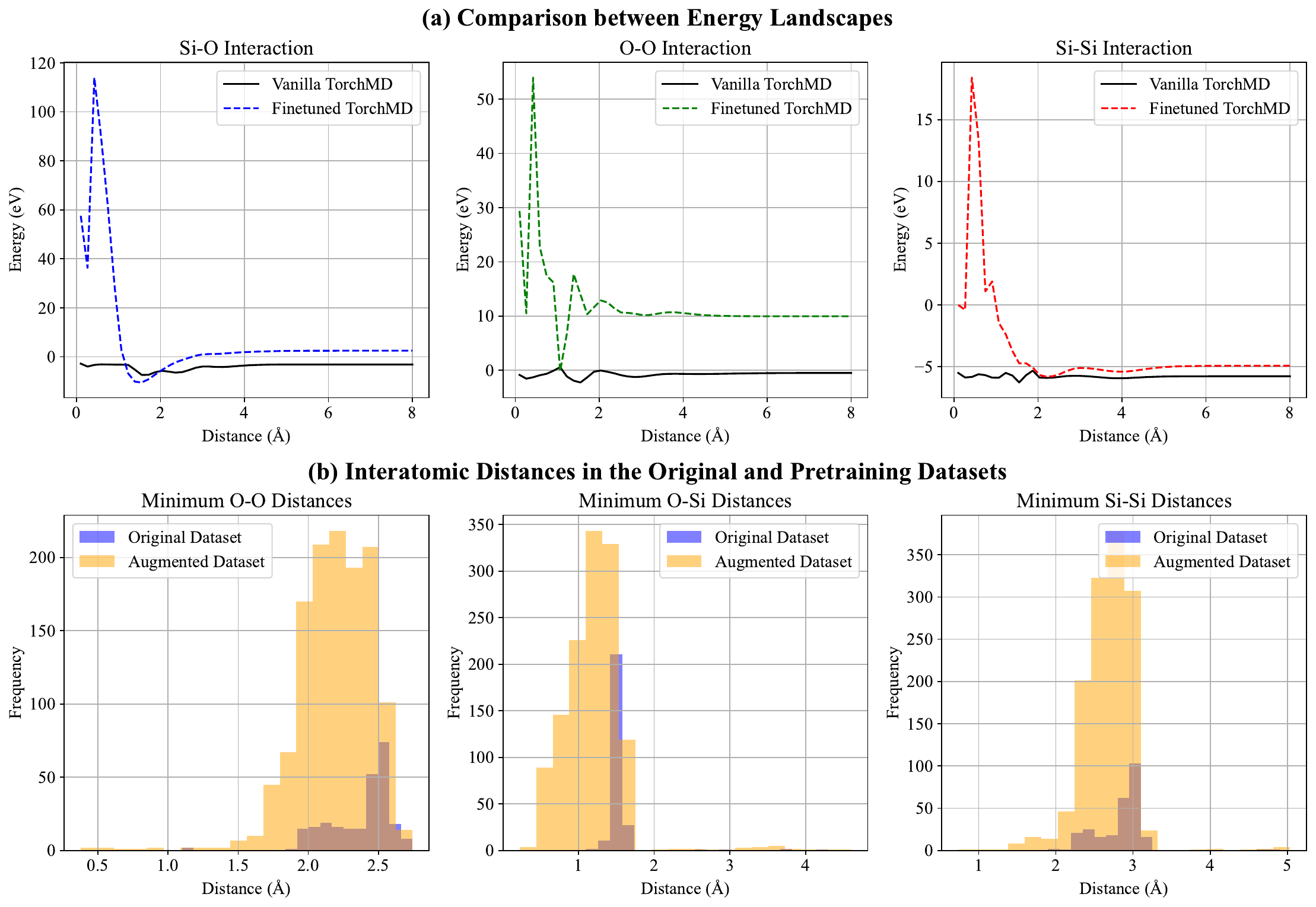}
\caption{Visualization of knowledge transfer from EAM potential to MLIP. (a) Finetuned model (dashed lines) versus baseline model (black solid lines) trained exclusively on DFT data. The finetuned MLIP features superior physical reasonableness for modeling the three key types of interactions in the Silica dataset, particularly in repulsive regions where the baseline model exhibits poor extrapolation. (b) Minimum interatomic distance distribution analysis: for each structure in the original and the pretraining datasets, we compute the minimum Si-Si, O-O, and Si-O distances and form distributions. The pretraining dataset (orange) significantly broadens the minimum distance coverage compared to the original silica dataset (blue) for all interaction pairs. The pretraining dataset contains compressed configurations that are underrepresented in quantum mechanical training sets and enables the MLIP to learn repulsive interactions during the pretraining stage.}
\label{fig:knowledge_transfer}
\end{figure}

The effectiveness of our physics-informed pretraining approach can be qualitatively demonstrated through a visualization of the knowledge transfer process from the EAM potential to the final finetuned MLIP, as illustrated in Figure \ref{fig:knowledge_transfer}. This analysis reveals how the sequential pretraining-finetuning protocol improves physicality of interatomic interactions across a wider configurational space.

We leverage the Silica dataset for demonstration purposes, as it contains a moderate number of interactions to be displayed. The knowledge transfer from EAM to MLIP occurs through structured exposure to physical energy landscapes during the pretraining phase. Figure \ref{fig:knowledge_transfer}(a) demonstrates the improved physicality of the finetuned model compared to a baseline model trained only on the original dataset. The finetuned model exhibits more physically reasonable potential curves, particularly in the repulsive regions where limited training data typically leads to poor extrapolation as shown by the energy landscape of the baseline MLIP. This improved behavior stems from the physics-informed initialization, which provides a chemically reasonable starting point for the optimization process.

A critical advantage of our approach lies in the systematic expansion of configurational coverage during pretraining. Figure \ref{fig:knowledge_transfer}(b) illustrates how the pretraining dataset (orange histograms) significantly broadens the interatomic distance distributions compared to the original quantum mechanical dataset (blue histograms). This expansion is evident for all three interaction types: O-O, O-Si, and Si-Si pairs. With the proposed perturbation schema, we systematic sample structures that include compressed configurations (with shorter interatomic distances) of structures, which are typically underrepresented in typical DFT training sets, as demonstrated by the histogram for the original silica dataset in Figure \ref{fig:knowledge_transfer}(b), but may be encountered during molecular dynamics simulations. This expanded coverage of structural configurations enables the MLIP to learn physical interpolation behavior across a broader range of structural configurations, the source of which is the EAM-provided physically consistent labels for these extended configurations. Consequently, the final finetuned MLIP exhibits enhanced stability and reduced unphysical behaviors during MD simulations.

\section{Discussion}

\subsection{Computational Overhead of the Proposed Method}

One of the key advantages of leveraging low-fidelity EAM data is that it avoids computationally expensive \textit{ab initio} computations for labeling the augmented dataset. Although training the EAM potential and the pretraining step introduce additional computational overhead, the superior efficiency of EAM-based labeling compared to DFT calculations makes the proposed approach computationally advantageous. Using the MPtrj Subset with TorchMD as an illustrative example, we provide timing estimates for each workflow component: training the EAM potential for 200 epochs requires 713.17 seconds ($\sim 12$ minutes), performing 25$\times$ data augmentation takes 73.20 seconds, and labeling the resulting 11,775 augmented structures with EAM requires 233.57 seconds ($\sim$ 4 minutes). The subsequent TorchMD pretraining on the augmented dataset takes 16 hours for 50 epochs, while the final 200-epoch finetuning on the original MPtrj (10\%) quantum-labeled dataset requires 2.44 hours. It should be noted that the TorchMD pretraining overhead would also be required if the augmented data were labeled using DFT calculations instead of EAM. In contrast, generating DFT labels for 11,775 additional structures (ranging from 11-104 atoms) would require substantial computational resources and time that far exceeds our EAM-based approach. The total additional overhead introduced by our method (EAM training + augmentation + EAM labeling + pretraining) amounts to approximately 17 hours compared to the baseline 2.44-hour training, which remains practically feasible while avoiding the prohibitive cost of extensive DFT calculations. All timing measurements were performed on a single NVIDIA A100 40GB GPU.

\subsection{Limitations and Future Area of Research}
While our physics-informed pretraining approach demonstrates clear advantages on the three selected datasets, several notable limitations of our approach should be discussed. The effectiveness of this method is significantly dependent on the accuracy and physical robustness of the underlying model used to label the pretraining data. The EAM potential used in this work performs poorly for problems with high elemental complexity; as a result we choose to evaluate our method for datasets containing four or less unique elements. In situations where the energy and force errors of the EAM potential are high, the benefits of pretraining are significantly reduced. 

Meanwhile, there is a growing body of research in developing foundational MLIPs which are trained on vast amounts of computational data. These include models such as SevenNet \cite{SevenNet}, MatterSim \cite{MatterSim}, ORB \cite{Orb}, EquiformerV2 \cite{Equiformer, barroso2024open, levine2025open}, and MACE \cite{MACE}. It is likely that with time, these class of models may serve as more effective teacher models than the classical empirical potentials used in this work. However, caution should still be taken to ensure these foundational MLIPs behave physically in extreme out-of-distribution scenarios as the more functionally constrained potentials. For example, recent systematic studies have revealed that current foundational models exhibit potential energy surface (PES) softening behavior, characterized by energy and force underprediction due to biased sampling of near-equilibrium atomic arrangements in training datasets \cite{PES-Softening}. Performance assessments by Focassio et al. \cite{UMLIP-Perm-Assess} also show that while these models excel in bulk predictions, they face challenges in extrapolating to out-of-distribution scenarios during surface energy calculations.

\section{Conclusion}\label{sec:conclusion}

In this study, we presented a physics-informed pretraining methodology that leverages an EAM potential to enhance the robustness and stability of MLIPs. Our approach employs a pretraining-finetuning pipeline wherein MLIPs are initially trained on EAM-labeled augmented datasets to acquire physical knowledge contained in the EAM potential before being finetuned on quantum mechanical reference data.

Comprehensive evaluation across three diverse material systems and three MLIP architectures demonstrates that our method consistently improves both prediction accuracy and molecular dynamics trajectory stability compared to the baseline model, SAM optimization and direct EAM-MLIP hybridization strategies. The pretraining methodology shows particular effectiveness in reducing unphysical behaviors during MD simulations, with substantial reductions in atom overlap incidents and Lindemann index violations across all tested systems. The visualization of the potential energy landscape of the finetuned MLIP and the analysis of force vector alignment reveals that our pretraining approach enables MLIPs to learn physics-consistent predictions that align well with underlying EAM profiles, which translates to improved capability of MLIPs in out-of-distribution scenarios.

While the current implementation is limited by the accuracy constraints of classical EAM potentials, this work establishes a viable framework for incorporating physics-based knowledge with low-fidelity data into MLIP training and demonstrates that the integration of classical physics knowledge through pretraining represents a promising direction for developing more robust and reliable MLIPs for computational materials science applications. Future developments alongside this direction may focus instead on incorporating more sophisticated foundational models as teacher networks to further enhance the accuracy and applicability of physics-informed MLIPs across broader chemical spaces.

\section{Data and Code Availability}

Data and model checkpoints supporting the results presented in this study are available at \url{https://huggingface.co/datasets/qzheng75/EAM-Physical-MLIP/tree/main}. The CGCNN and TorchMD checkpoints can be used with the MatDeepLearn package \cite{MatDeepLearn}, while the M3GNet checkpoints are provided for integration with the MatterTune package \cite{MatterTune}. The implementation of the EAM potential under the MatDeepLearn framework is available at \url{https://github.com/Fung-Lab/MatDeepLearn_dev/blob/eam_stable_md/matdeeplearn/models/eam_interaction.py} and a sample implementation of the two-step data augmentation process can be found at \url{https://github.com/Fung-Lab/MatDeepLearn_dev/blob/eam_stable_md/scripts/augmentation.py}.

\section{Acknowledgements}




\clearpage
\bibliography{refs}
\bibliographystyle{naturemag}
\clearpage

\appendix
\counterwithin{figure}{section}
\counterwithin{table}{section}
\counterwithin*{equation}{section}
\renewcommand\theequation{\thesection\arabic{equation}}

\begin{titlepage}
  \centering
  \LARGE \textsc{Supplementary Information} \par
  \let\endtitlepage\relax
\end{titlepage}

\section{Detailed experimental setup for constructing the pretraining dataset}\label{sec:appendix-a}

The comprehensive experimental configurations are summarized in Table \ref{tab:experimental_config}. The notation $[a, b)$ represents a linearly spaced sequence of perturbation strengths generated using numpy's linspace function, where $a$ is the starting value, $b$ is the endpoint (exclusive), and the number of points equals the augmentation factor $n_{aug}$. For instance, $[a, b)$ with five augmentations generates the sequence $\{1.0,0.8,0.6,0.4,0.2\}$ \r{A}. Each augmentation round applies our dual-mode perturbation algorithm with the corresponding strength values, resulting in a pretraining dataset that is $n_{aug}$ times larger than the original quantum mechanical reference data.

\begin{table}[htbp]
\centering
\small
\begin{tabular}{|c|c|c|c|c|c|c|c|}
\hline
\textbf{Dataset} & \textbf{MLIP} & \textbf{Usage} & \textbf{$n$ aug} & \textbf{$n$ pairs} & \textbf{Pull} & \textbf{$n$ sites} & \textbf{Perturb} \\
\hline
& CGCNN & 100\% & 5 & 4 & [1.0, 0.0) & 5 & [1.0, 0.0) \\
Silica & TorchMD & 10\% & 5 & 4 & [1.0, 0.0) & 5 & [1.0, 0.0) \\
& M3GNet & 10\% & 5 & 2 & [1.0, 0.0) & 5 & [1.0, 0.0) \\
\hline
& CGCNN & 100\% & 5 & 2 & [1.0, 0.0) & 5 & [1.0, 0.0) \\
Phosphorus & TorchMD & 5\% & 20 & 2 & [2.0, 0.0) & 5 & [2.0, 0.0) \\
& M3GNet & 5\% & 5 & 2 & [1.0, 0.0) & 5 & [0.5, 0.0) \\
\hline
& CGCNN & 100\% & 10 & 2 & [0.5, 0.0) & 5 & [0.25, 0.0) \\
MPtrj Subset & TorchMD & 10\% & 25 & 2 & [1.0, 0.0) & 5 & [0.5, 0.0) \\
& M3GNet & 10\% & 25 & 2 & [1.0, 0.0) & 5 & [0.5, 0.0) \\
\hline
\end{tabular}
\caption{Experimental configuration parameters for physics-informed pretraining. Usage indicates the fraction of original dataset used for training; $n$ aug is the number of augmentation rounds applied; $n$ pairs and $n$ sites are the number of close atomic pairs identified for pulling and the number of sites for random perturbation, respectively; Pull and Perturb specify the strength ranges for atomic pulling and random perturbation, respectively.}
\label{tab:experimental_config}
\end{table}

\section{Comparison with Direct EAM Integration}

\begin{table}[htbp]
\centering
\renewcommand{\arraystretch}{1.3}
\begin{tabular}{|c|c|c|c|c|}
\hline
\textbf{Dataset} & \textbf{MD Setting} & \textbf{Setup} & \textbf{Atom Overlap} & \textbf{Lindemann Index} \\
\hline
\multirow{2}{*}{Silica (full)} & NVT (2000K, 2fs) & \multirow{2}{*}{CGCNN + EAM} & 3 & 1  \\
\cline{2-2}\cline{4-5}
& NPT (2000K, 2fs) & & 6 & 3 \\
\hline
\multirow{2}{*}{Silica (10\%)} & NVT (2000K, 2fs) & \multirow{2}{*}{TorchMD + EAM} & 2 & 3 \\
\cline{2-2}\cline{4-5}
& NPT (2000K, 2fs) & & 9 & 8 \\
\hline
\multirow{1}{*}{Phosphorous (full)} & NVT (1250K, 1fs) & \multirow{1}{*}{CGCNN + EAM} & 39 & 67  \\
\hline
\multirow{1}{*}{Phosphorous (10\%)} & NVT (1250K, 1fs) & \multirow{1}{*}{TorchMD + EAM} & 11 & 79  \\
\hline
\multirow{2}{*}{MPtrj Subset (full)} & NVT (500K, 1fs) & \multirow{2}{*}{CGCNN + EAM} & 100 & 0 \\
\cline{2-2}\cline{4-5}
& NPT (250K, 1fs) & & 47 & 47 \\
\hline
\multirow{2}{*}{MPtrj Subset (10\%)} & NVT (1000K, 1fs) & \multirow{2}{*}{TorchMD + EAM} & 88 & 100  \\
\cline{2-2}\cline{4-5}
& NPT (750K, 1fs) & & 81 & 100 \\
\hline
\end{tabular}
\caption{Trajectory stability assessment for hybrid MLIPs across diverse datasets and simulation conditions. All molecular dynamics simulations were conducted for 5000 timesteps on 100 randomly sampled structures from each dataset's test set. Values represent the number of trajectories (out of 100) that exhibit unphysical behavior as detected by atom overlap and Lindemann index criteria. NVT and NPT simulations are performed with Langevin and Berendsen dynamics, respectively, as introduced in section \ref{sec:MD_intro}, and implemented with the Atomic Simulation Environment (ASE) package.}
\label{tab:hybrid_md_results}
\end{table}

At the early stage of our research, we investigated a seemingly easier approach to introduce chemical knowledge from traditional interatomic potentials by directly incorporating EAM-predicted energies as an additive term to the MLIP output, following the regularization strategy proposed in \cite{SN-Reg}. In this hybrid architecture, we first train the EAM potential independently, then jointly optimize both the EAM parameters and MLIP weights on quantum mechanical reference data. The total energy prediction combines contributions from both the MLIP and EAM components: $E_{total} = E_{MLIP} + E_{EAM}$. However, our evaluation across multiple datasets and MD simulation demonstrates that this hybrid approach yields inferior performance compared to our proposed sequential pretraining-finetuning methodology, as evidenced by both prediction accuracy metrics (Table \ref{tab:hybrid_MLIP_EFS}) and trajectory stability assessments (Table \ref{tab:hybrid_md_results}).

The comprehensive evaluation across three diverse datasets and multiple simulation conditions demonstrates the clear superiority of our proposed physics-informed pretraining methodology over the direct EAM-MLIP combination approach. The trajectory stability assessment (Table \ref{tab:md_results}) reveals the fundamental limitations of direct combination methods: the hybrid methods consistently exhibit substantially higher rates of trajectory instability across most tested systems compared to the proposed pretraining approach. These results demonstrate that sequential pretraining with EAM-derived knowledge provides superior physics-informed initialization that enhances both prediction accuracy and long-term simulation stability, whereas direct combination approaches fail to effectively leverage empirical potential knowledge and may harm trajectory physicality.

\begin{table}[htbp]
\centering
\renewcommand{\arraystretch}{1.3}
\begin{tabular}{|c|c|c|c|c|c|}
\hline
\textbf{Dataset} & \textbf{Setup} & \textbf{E Error} & \textbf{F Error} & \textbf{S Error} & \textbf{Total Error} \\
\hline
Silica & CGCNN + EAM & 2.973 & 0.193 & 0.00134 & 9.754 \\
\hline
Silica (10\%) & TorchMD + EAM & 2.490 & 0.116 & 0.00111 & 5.875 \\
\hline
Phosphorous & CGCNN + EAM & 8.476 & 0.264 & / & 13.280 \\
\hline
Phosphorous (10\%) & TorchMD + EAM & 8.396 & 0.208 & / & 10.489 \\
\hline
MPtrj Subset & CGCNN + EAM & 1.106 & 0.090 & 0.00188 & 4.611 \\
\hline
MPtrj Subset (10\%) & TorchMD + EAM & 2.558 & 0.086 & 0.00173 & 4.392 \\
\hline
\end{tabular}
\caption{Prediction accuracy comparison for hybrid MLIPs combining MLIPs with EAM potentials. Energy errors are reported as mean absolute error (MAE) in eV, force errors in eV/\r{A}, and stress errors in GPa. The total error represents a weighted combination of energy, force, and stress MAEs using weights 0.01:50:50, respectively. Dataset percentages indicate the fraction of available training data used for model development.}
\label{tab:hybrid_MLIP_EFS}
\end{table}

\begin{figure}[h!]
\centering
\includegraphics[width=1.0\textwidth]{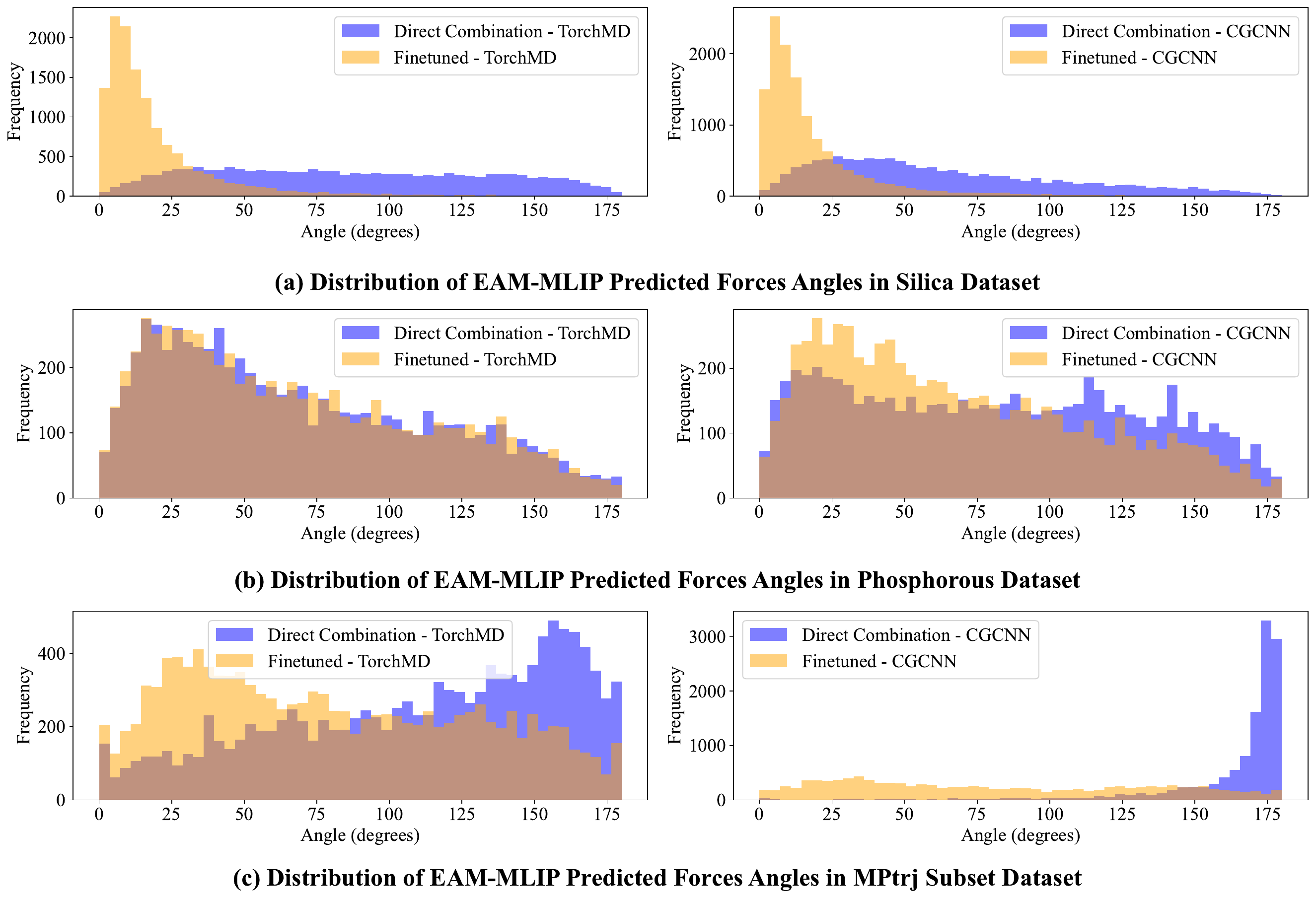}
\caption{Angular distribution analysis of force predictions comparing direct EAM combination versus pretraining approaches across three datasets. The histograms show the distribution of angles between MLIP-predicted and EAM-predicted force vectors for individual atoms, where $0^{\circ}$ indicates perfect alignment and $180^{\circ}$ represents complete opposition. Orange distributions represent models trained with our proposed pretraining-finetuning methodology, while blue distributions show results from direct EAM-MLIP combination approaches.}
\label{fig:angle_distribution}
\end{figure}

To explore the inferior performance of the direct combination approach, we investigated how the MLIP aligns with the EAM potential in the direct combination method and the proposed pretraining-finetuning method. The angular distribution analysis reveals a fundamental incompatibility between direct EAM-MLIP combination approaches and highlights the effectiveness of our proposed pretraining methodology in achieving coherent physics-informed learning. For the direct combination approach, we performed inference with the hybrid model on test structures from each dataset and decomposed the total force predictions to separately extract the EAM and MLIP force contributions for each atom, then computed the angles between these concurrent force vectors using the dot product $\theta = \arccos\left(\frac{\mathbf{F}_{EAM} \cdot \mathbf{F}_{MLIP}}{|\mathbf{F}_{EAM}||\mathbf{F}_{MLIP}|}\right)$. The results, as shown in Figure \ref{fig:angle_distribution}, demonstrate that these two components frequently ``fight against" each other, producing conflicting force directions that manifest as broad angular distributions with substantial density at large angles ($90-180^{\circ}$), particularly evident in the Silica and MPtrj datasets where direct combination methods exhibit significant force misalignment. This may explain the especially inferior MD stability performance of the hybrid MLIP in these two datasets. In contrast, for our pretraining approach, we computed angles between force predictions from the finetuned MLIP and the EAM potential used for pretraining dataset generation when both models independently performed inference on the test structures. The angular distributions are more concentrated near small angles, which indicates a more consistent force alignment across all tested systems and demonstrates that our ``soft knowledge injection" strategy through sequential pretraining enables the MLIP to internalize and harmonize with EAM physics principles, rather than competing against them. The pretraining methodology thus facilitates smoother EAM integration by allowing the MLIP to learn representations that respect rather than contradict the underlying empirical potential, resulting in more stable and physically consistent predictions during inference.

\section{Hyperparameters for MLIP architectures}

In this section, we presented detailed hyperparameters for the selected MLIP architectures. The configurations for each architecture are transferrable to all three datasets.

\textbf{CGCNN}: The CGCNN architecture employs a 4-layer graph convolutional network with hidden dimensions of 100 in the convolutional layers, followed by a post-GNN MLP with hidden dimension of 150. The model includes 1 pre-fully connected layer and 3 post-fully connected layers and utilizes global addition pooling applied late in the architecture with SiLU activation functions.

\textbf{TorchMD}: The TorchMD architecture employs an 8-layer graph neural network with 128 hidden channels and 50 radial basis functions, incorporating 8 attention heads with both SiLU activation and attention activation functions. The model includes 2 post-processing layers with 128 post-hidden channels and utilizes global addition pooling applied late in the architecture with additive aggregation.

\textbf{M3GNet}: The selected M3GNet architecture has the same configuration as the backbone model of the foundational model MatterSim-v1.0.0-1M. Details about the architecture can be found at \cite{MatterSim}.

\end{document}

%% file: math_commands.tex
\usepackage{amsmath,amsfonts,bm}

\def\eqref#1{equation~\ref{#1}}

\def\1{\bm{1}}

\DeclareMathAlphabet{\mathsfit}{\encodingdefault}{\sfdefault}{m}{sl}
\SetMathAlphabet{\mathsfit}{bold}{\encodingdefault}{\sfdefault}{bx}{n}

